\documentclass[reprint,aps,pra,amsmath,amssymb,tightenlines,epsfig,floatfix, twocolumn, longbibliography]{revtex4-2}
\usepackage{natbib}
\usepackage[export]{adjustbox}
\usepackage{graphicx}
\usepackage{dcolumn}
\usepackage{bm}
\usepackage{physics}
\usepackage{amsfonts}
\usepackage{xspace}
\usepackage{color}
\usepackage{xcolor}
\usepackage{epstopdf}
\usepackage{multirow}
\usepackage{hyperref}

\begin{document}
\newcommand{\psihat}{\ensuremath{\hat{\psi}}\xspace}
\newcommand{\psihatd}{\ensuremath{\hat{\psi}^{\dagger}}\xspace}
\newcommand{\ahat}{\ensuremath{\hat{a}}\xspace}
\newcommand{\Ham}{\ensuremath{\mathcal{H}}\xspace}
\newcommand{\ahatd}{\ensuremath{\hat{a}^{\dagger}}\xspace}
\newcommand{\bhat}{\ensuremath{\hat{b}}\xspace}
\newcommand{\bhatd}{\ensuremath{\hat{b}^{\dagger}}\xspace}
\newcommand{\boldr}{\ensuremath{\mathbf{r}}\xspace}
\newcommand{\dr}{\ensuremath{\,d^3\mathbf{r}}\xspace}
\newcommand{\dk}{\ensuremath{\,d^3\mathbf{k}}\xspace}
\newcommand{\etal}{\emph{et al.\/}\xspace}
\newcommand{\ie}{i.e.}
\newcommand{\eq}[1]{Eq.~(\ref{#1})\xspace}
\newcommand{\fig}[1]{Fig.~\ref{#1}\xspace}
\newcommand{\proj}[2]{\left| #1 \rangle\langle #2\right| \xspace}
\newcommand{\Qhat}{\ensuremath{\hat{Q}}\xspace}
\newcommand{\Qhatd}{\ensuremath{\hat{Q}^\dag}\xspace}
\newcommand{\phihatd}{\ensuremath{\hat{\phi}^{\dagger}}\xspace}
\newcommand{\phihat}{\ensuremath{\hat{\phi}}\xspace}
\newcommand{\boldk}{\ensuremath{\mathbf{k}}\xspace}
\newcommand{\boldp}{\ensuremath{\mathbf{p}}\xspace}
\newcommand{\boldsigma}{\ensuremath{\boldsymbol\sigma}\xspace}
\newcommand{\boldalpha}{\ensuremath{\boldsymbol\alpha}\xspace}
\newcommand{\parti}[2]{\frac{ \partial #1}{\partial #2} \xspace}
 \newcommand{\vs}[1]{\ensuremath{\boldsymbol{#1}}\xspace}
\renewcommand{\v}[1]{\ensuremath{\mathbf{#1}}\xspace}
\newcommand{\Psihat}{\ensuremath{\hat{\Psi}}\xspace}
\newcommand{\Psihatd}{\ensuremath{\hat{\Psi}^{\dagger}}\xspace}
\newcommand{\Vhatd}{\ensuremath{\hat{V}^{\dagger}}\xspace}
\newcommand{\Xhat}{\ensuremath{\hat{X}}\xspace}
\newcommand{\Xhatd}{\ensuremath{\hat{X}^{\dag}}\xspace}
\newcommand{\Yhat}{\ensuremath{\hat{Y}}\xspace}
\newcommand{\Jhat}{\ensuremath{\hat{J}}\xspace}
\newcommand{\Yhatd}{\ensuremath{\hat{Y}^{\dag}}\xspace}
\newcommand{\jhat}{\ensuremath{\hat{J}}\xspace}
\newcommand{\lhat}{\ensuremath{\hat{L}}\xspace}
\newcommand{\Nhat}{\ensuremath{\hat{N}}\xspace}
\newcommand{\rhohat}{\ensuremath{\hat{\rho}}\xspace}
\newcommand{\ddt}{\ensuremath{\frac{d}{dt}}\xspace}
\newcommand{\nset}{\ensuremath{n_1, n_2,\dots, n_k}\xspace}
\newcommand{\notes}[1]{{\color{blue}#1}}
\newcommand{\sah}[1]{\textcolor{red}{#1}}


\title{A hybrid method of generating spin-squeezed states for quantum-enhanced atom interferometry}
\author{Liam A. Fuderer}
\author{Joseph J. Hope}
\author{Simon A. Haine} 
\email{simon.a.haine@gmail.com}
\affiliation{%
Department of Quantum Science and Technology, Research School of Physics, The Australian National University, Canberra, Australia}
\date{\today}

\begin{abstract}
We introduce a new spin-squeezing technique that is a hybrid of two well established spin-squeezing techniques, quantum nondemolition measurement (QND) and one-axis twisting (OAT). This hybrid method aims to improve spin-squeezing over what is currently achievable using QND and OAT. In practical situations, the strength of both the QND and OAT interactions is limited. We found that in these situations, the hybrid scheme performed considerably better than either OAT or QND used in isolation. As QND and OAT have both been realised experimentally, this technique could be implemented in current atom interferometry setups with only minor modifications to the experiment.

\end{abstract}

\maketitle


\begin{figure*}[!t] 

    \includegraphics[width=0.9\textwidth]{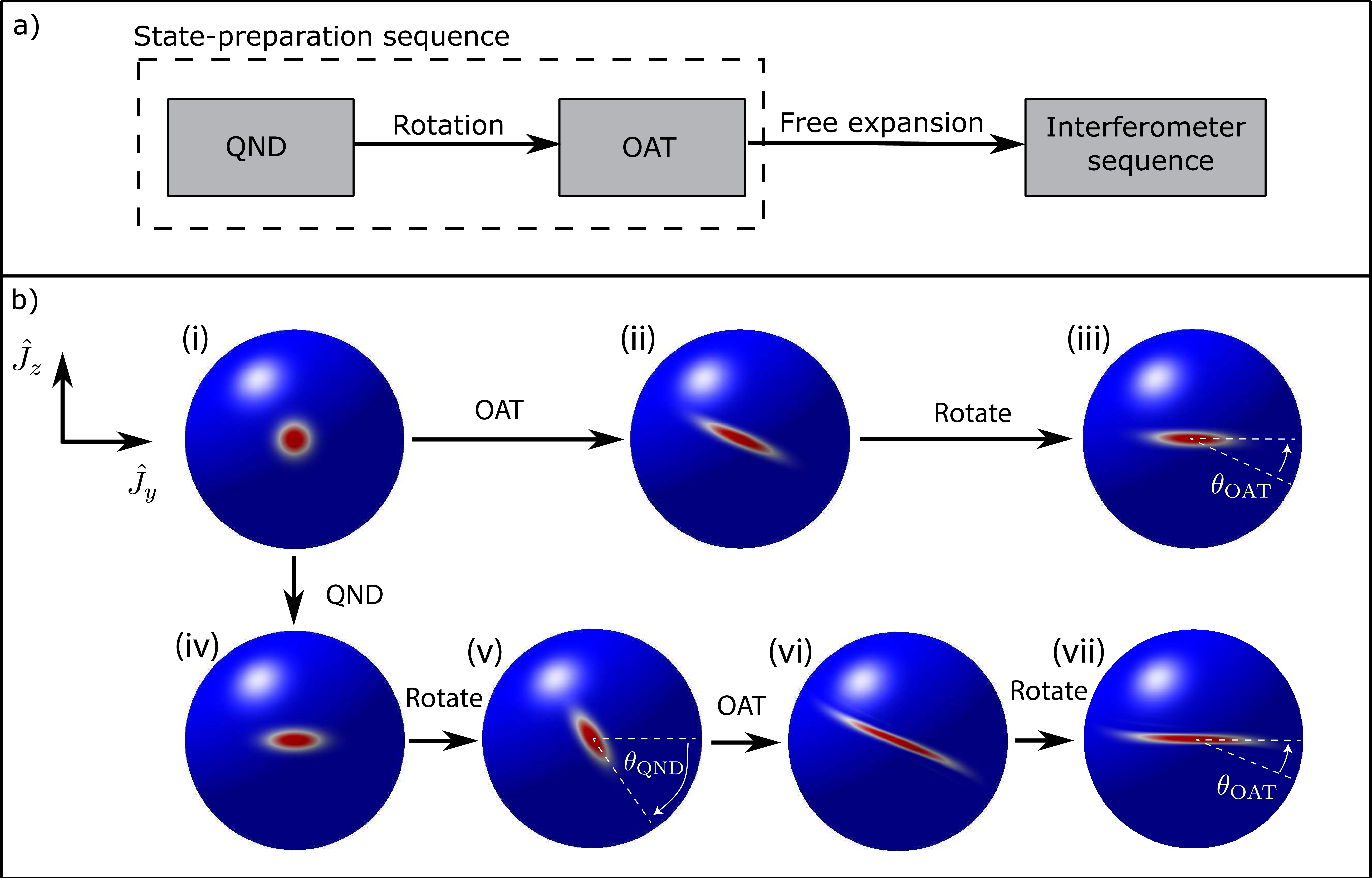}
    \caption{\label{fig:Bloch}a) Hybrid method schematic. The model is a modification to the current OAT state-preparation sequence. The output state from the hybrid model is used as the initial state for the interferometer sequence. b) Bloch sphere representation of OAT (top row) and the proposed hybrid method (bottom row) for an initial CSS. In OAT, the initial CSS (i) undergoes nonlinear shearing creating a state with reduced variance in some direction (ii). This state is then rotated about the $J_x$ axis by an amount $\theta_\mathrm{OAT}$ in order to create a state with reduced variance in the $J_z$ axis (iii). In the hybrid method, the initial state first undergoes QND squeezing, creating a state with reduced variance in the $J_z$ axis (iv). This state is then rotated about the $J_x$ axis by an amount $\theta_\mathrm{QND}$ (v), such that it undergoes more rapid shearing under OAT dynamics (vi). This state is then rotated by amount $\theta_\mathrm{OAT}$ to reduce the variance in the $J_z$ direction. When the degree of OAT or QND interaction is limited, the hybrid scheme can produce better spin squeezing than either OAT or QND used in isolation. }
\end{figure*}

\section{\label{sec:intro}Introduction}
Atom interferometry is capable of providing state-of-the-art measurements of gravitational fields \cite{Peters:1999, Peters:2001, Hu:2013, Hauth:2013, Menoret:2018, Bongs:2019, Kritsotakis:2018}, gravitational gradients \cite{Snadden:1998, McGuirk:2002, Sorrentino:2014, Biedermann:2015, DAmico:2016, Asenbaum:2017}, and magnetic fields \cite{Vengalattore:2007}, with future applications such as minerals exploration \cite{Evstifeev:2017, Geiger:2020}, hydrology \cite{Canuel:2018}, inertial navigation \cite{Jekeli:2005, Battelier:2016, Cheiney:2018, Wang:2023}, and possible tests of candidate theories of quantum gravity \cite{Dimopoulos:2007, Haine:2021, Tino:2021}. There is considerable recent interest in the use of quantum entanglement in atom interferometry \cite{Pezze_review:2018, Szigeti:2021, Colombo:2022}, which would improve the precision, measurement rate, and reduce the overall size of these devices \cite{Szigeti:2021}. Two of the most promising routes to generating useful entanglement are spin-squeezing via One-Axis Twisting (OAT) \cite{Kitagawa:1993, Esteve:2008, Gross:2010, Riedel:2010}, or via Quantum non-demolition (QND) measurements \cite{Appel:2009, Takano:2009, Louchet-Chauvet:2010, Koschorreck:2010, Schleier-Smith:2010, Leroux:2010, Sewell:2012}. These methods have been used in proof of principle experiments \cite{Hosten:2016, Greve:2022}, but have not yet found utility in a state-of-the-art inertial sensors. In particular, typical atom interferometry experiments with ultra-cold atoms have only modest optical densities, and permit only modest levels of squeezing via QND \cite{Kritsotakis:2021}. While the use of an optical cavity can significantly improve the amount of achievable squeezing \cite{Schleier-Smith:2010, Leroux:2010, Hosten:2016, Greve:2022}, this adds considerable experimental overhead, increasing the size, weight, and complexity of the experiment \cite{Battelier:2016}.

A recent proposal showed that in a freely expanding Bose-Einstein condensate (BEC), strong atom-atom interactions can generate substantial spin-squeezing via OAT between two momentum modes without degrading mode overlap or causing significant phase diffusion, and could potentially allow for a high precision, spin-squeezed gravimetry measurement \cite{Szigeti:2020}. However, only a modest level of OAT interactions are achievable via this method, as the interactions are quickly reduced due to the expansion of the atomic clouds. Here, we present a hybrid scheme that utilises both QND and OAT. In particular, by combining both schemes, we can achieve levels of squeezing significantly higher than either scheme on their own. When restricting ourselves to the small levels of QND interaction that free-space QND permits, and the weak OAT interaction that would be generated via the scheme presented in Szigeti \etal \cite{Szigeti:2020}, we show that combining these schemes can give significantly better levels of squeezing than either in isolation. Furthermore, these schemes are entirely compatible, and can be implemented together without compromising the performance of either one.

\section{\label{sec:hybrid} Combining QND and OAT to improve spin-squeezing}
We will begin by describing the hybrid scheme, and then going into the details of each element, specifically, the QND and OAT interactions. Assuming a BEC of $N$ atoms with two hyperfine states $|1\rangle$ and $|2\rangle$, we introduce the pseudo-spin operators $\hat{J}_k = \frac{1}{2}(\hat{a}_1^\dagger ~\hat{a}_2^\dagger) \sigma_k (\hat{a}_1~ \hat{a}_2)^T$, where $\sigma_k$ is the $k$th Pauli matrix, and $\hat{a}_1$ and $\hat{a}_2$ are the annihilation operators for atomic states $|1\rangle$ and $|2\rangle$ respectively. These two states may also carry an associated momentum difference, such as is the case in an atom interferometer used to measure gravity. When used as the input to a Mach-Zehnder interferometer, the achievable phase-sensitivity is
\begin{align}
\Delta \phi = \frac{\xi}{\sqrt{N}}
\end{align}
where
\begin{align}
\xi = \sqrt{N}\frac{\sqrt{\mathrm{Var}(\hat{J}_z)}}{| \langle \hat{J}_x\rangle |}
\end{align}
where $\xi$ is the Wineland spin-squeezing parameter \cite{Wineland:1992}, with $\xi <1$ indicating spin-squeezing. The hybrid scheme involves first applying the QND interaction, and then using the OAT interaction to further enhance the squeezing (see figure (\ref{fig:Bloch})). Specifically, we initially prepare our system with all atoms in state $|1\rangle$ (maximum $\hat{J}_z$ eigenstate), before applying a beamsplitter pulse which puts each atom in an equal coherent superposition of $|1\rangle$ and $|2\rangle$. This is a rotation about the $\hat{J}_y$ axis by $\pi/2$, creating a coherent spin-state (CSS) on the equator of the Bloch sphere, or a maximal $\hat{J}_x$ eigenstate. The QND interaction reduces the uncertainty in the $\hat{J}_z$ axis, while increasing fluctuations in the $\hat{J}_y$ axis. The state is then rotated by an angle $\theta_\mathrm{QND}$ about the $\hat{J}_x$ axis, before the OAT interaction is applied, which causes a non-linear `shearing' of the state. Rotating the QND state before the OAT interaction increases the variance in $\hat{J}_z$, which causes the state to shear faster under the OAT interaction, ultimately increasing the amount of spin-squeezing achievable. The value of $\theta_\mathrm{QND}$ that optimises the spin-squeezing parameter depends on both the amount of QND interaction before, and OAT interaction after the rotation. The QND interaction is achieved via an optical coupling, and can occur on time-scales much faster than the atomic dynamics, while the OAT interaction is achieved via utilisation of the interatomic interactions, and typically takes several milliseconds to achieve significant shearing. As these two interactions utilise different resources, they are entirely compatible and can be applied sequentially as described above. As QND and OAT have both been realised experimentally, this technique could be implemented in current atom interferometry setups with only minor modifications to the experiment. Specifically, as described in \cite{Szigeti:2020}, OAT can be achieved by replacing the free-expansion time with two additional beamsplitter pulses, implemented via the same laser system as the interferometric pulses. QND is achieved by using an off-resonant laser (or a pair of lasers, one for each hyperfine state) to perform state-dependent number estimation of the sample after an initial additional beamsplitter pulse. 

We will describe the dynamics of this scheme quantitatively in section \ref{sec:hybrid}. We will now briefly review the principles and performance of the OAT and QND interactions individually, before assessing the performance of the hybrid scheme. 

\subsection{QND-Squeezing}
By illuminating the atomic sample with a laser detuned from some excited state $|e\rangle$, the population in each of the hyperfine states $|a_1\rangle$ and $|a_2\rangle$ is imprinted on the phase of the light. Consequently, a measurement of the phase allows one to infer information about the population difference, collapsing the atomic state into a spin-squeezed state (SSS), with reduced variance in $\hat{J}_z$  \cite{Kuzmich:1998, Pezze_review:2018}. As the collapse is random, feedback (via a rotation around the $\hat{J}_y$ axis of magnitude proportional to the measurement result) is used to re-center the state on the equator, such that $\langle \hat{J}_z\rangle = 0$. The amount of spin-squeezing depends on the strength of the atom-light entanglement. QND can be achieved using monochromatic (one-colour) \cite{Oblak:2005}, or dichromatic (two-colour) laser light \cite{Saffman:2009, Kritsotakis:2021}. One-colour QND is susceptible to dephasing of the atoms due to the inhomogeneous spatial profile of the laser, which degrades spin-squeezing \cite{Windpassinger:2008}. Two-colour QND rectifies this issue and suppresses experimental noise, such as vibrations in mirror positions, to first order. Assuming the laser is spatially homogeneous, both methods can be shown to have the same spin-squeezing. For simplicity, we will therefore proceed with a one-colour QND model without loss of generality, although a two-colour scheme may be favourable for experimental implementation.

\begin{figure}[h!]
    \includegraphics[width=\columnwidth]{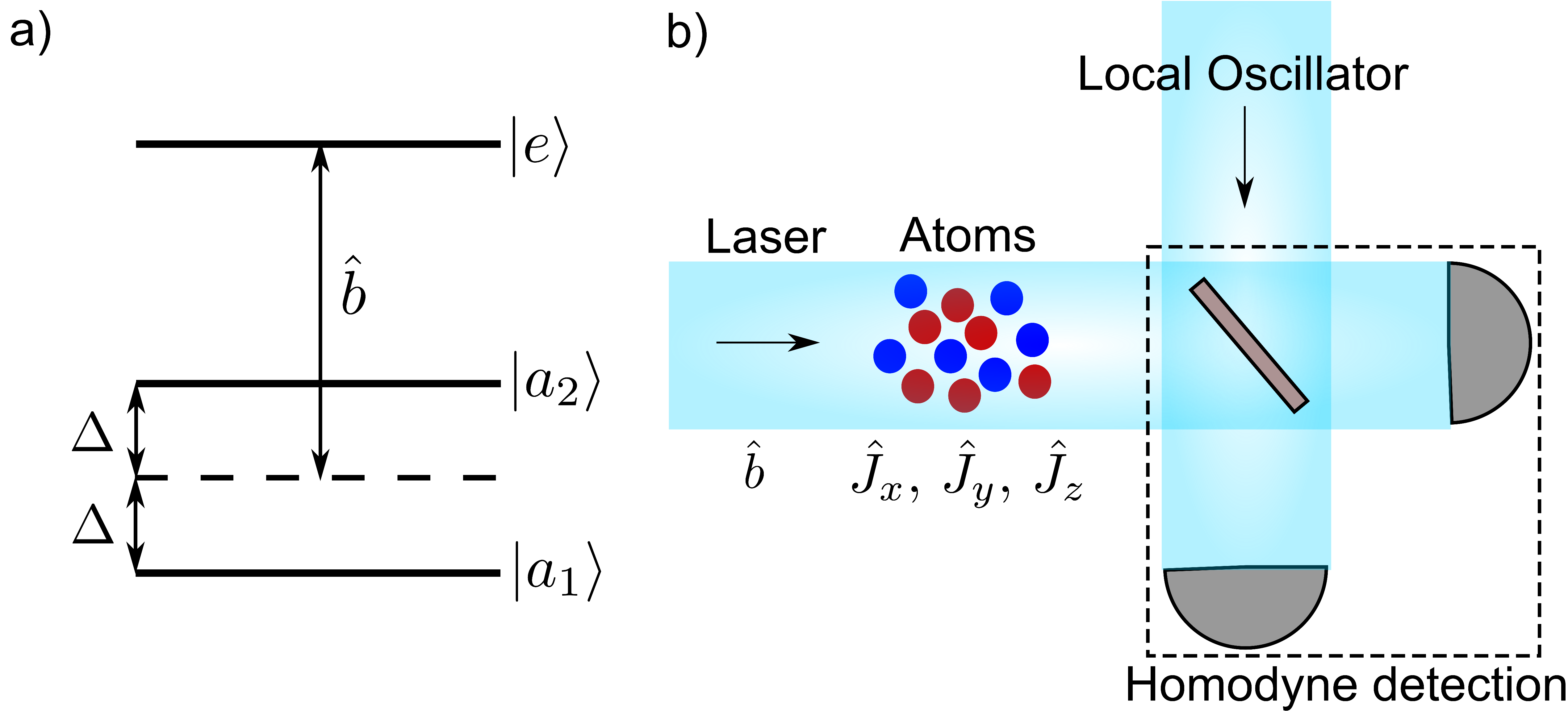}
    \caption{\label{fig:QND} a) Energy-level diagram of quantum non-demolition measurements. The atoms are illuminated by a laser (annihilation operator \(\hat{b}\)) equally detuned from the two atomic levels, \(\ket{a_1}\) and \(\ket{a_2}\), to an excited state \(\ket{e}\). This imprints the difference in the number of atoms in each state, \(\hat{J}_z\), into the phase of the laser. b) Schematic of experimental quantum non-demolition measurement in free-space. After the laser passes through the atoms, the phase is measured by a homodyne detector, allowing for an inference of \(\hat{J}_z\) to be made. Such an inference reduces the variance in \(\hat{J}_z\) and thus creates a SSS.}
\end{figure}

A schematic of the atom-light interaction is shown in Figure \ref{fig:QND}. A single laser is detuned from resonance by an amount $(-)\Delta$ for the $|a_{1(2)}\rangle \rightarrow |e\rangle$ transition. Adiabatically eliminating the excited state \cite{Kuzmich:1998} $|e\rangle$ gives the effective hamiltonian
 \begin{eqnarray}
    \hat{H}_{\mathrm{QND}} &=& -\hbar \chi_{\mathrm{QND}} \hat{J}_z \hat{b}^\dagger \hat{b}, \label{HQND}
\end{eqnarray}
where $\hat{b}$ is the annihilation operator for a pulse of light of duration $t_p$, and $\chi_{\mathrm{QND}}$ is the atom-light interaction strength, which for the detuning $\Delta$ shown in figure \ref{fig:QND} is
\begin{eqnarray}
	\chi_{\mathrm{QND}} = \frac{2\sigma_0 \Gamma}{A\Delta t_p}.
\end{eqnarray}
Here, $\sigma_0$ is the resonant scattering cross-section, $\Gamma$ is the transition spontaneous emission rate, $A$ is the cross-sectional area of the laser incident on the atomic sample. 

\begin{figure}[h!]
    \includegraphics[width=\columnwidth]{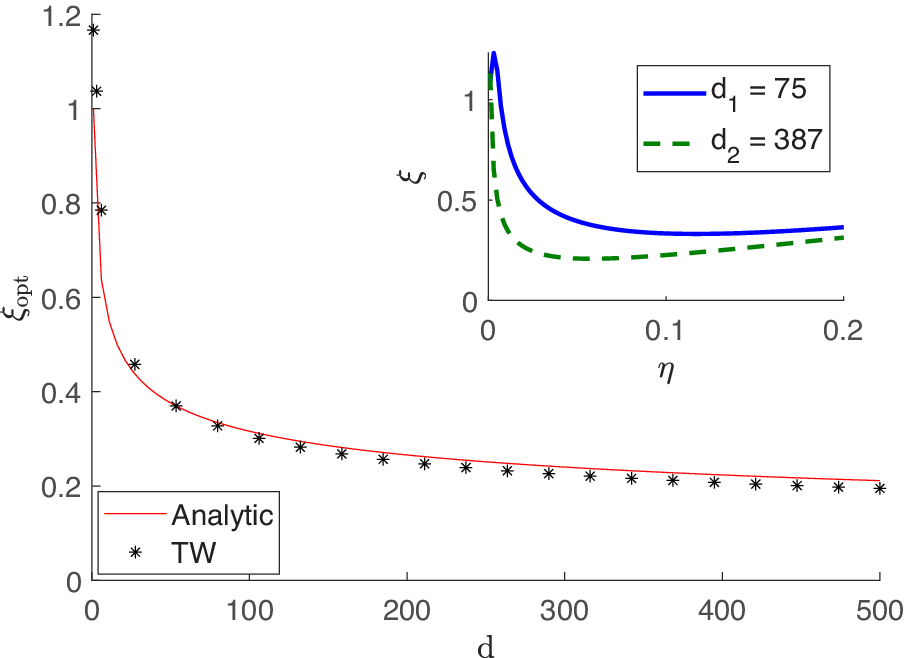}
    \caption{\label{fig:QNDScaling} Optimum spin-squeezing parameter $\xi_\mathrm{opt}$ from QND on the D1 line of a system of $10^5$ $^{87}$Rb atoms, calculated via Truncated Wigner (black stars), compared to analytic scaling $\xi_\mathrm{opt} = d^{-1/4}$ (red solid line).  Inset figure: Scaling of the spin-squeezing parameter, $\xi$, with the loss fraction $\eta$ for a fixed resonant optical depth. Outer figure: The optimal spin-squeezing parameter as a function of optical depth, found by minimisation of $\xi$ with respect to $\eta$, for $d=d_1 = 75$ (blue solid line) and $d=d_2 =387$ (green dashed line) . $\xi_{\mathrm{opt}}$ scales as $d^{-1/4}$.}
\end{figure}

The degree of spin-squeezing is well characterised by the resonant optical depth, \(d\), and the inelastic scattering rate integrated over the pulse duration, \(\eta\), which can be written in terms of the system parameters as
\begin{eqnarray}
    d &=& \frac{\sigma_0 N}{A}, \\
    \text{and }\eta &=& \frac{2\sigma_0}{A} \left(\frac{\Gamma}{\Delta}\right)^2 N_p, \label{etadef}
\end{eqnarray}
where \(N_p\) is the total photon number. Due to the narrow momentum linewidth of the Bose-Einstein condensate, any spontaneous emissions causes the atom to be scattered into a distinguishable momentum state, and is therefore lost from the interference measurement \cite{Kritsotakis:2021}. We treat these spontaneous emission events by coupling vacuum noise fluctuations into the atomic operators \cite{Echaniz:2005}.

For a fixed resonant optical depth, there is a trade-off between the amount of information inferred about the population difference and information lost due to spontaneous emission, as seen in the inset of Figure \ref{fig:QNDScaling}.  Optimising over the loss fraction gives \cite{Hammerer:2004, Saffman:2009, Appel:2009} an optimum squeezing of
\begin{equation}
\xi_{\mathrm{opt}} \approx d^{-1/4} .
\end{equation}
For Bose-condensed atoms, the atomic density is limited by three-body recombination, which ultimately limits the achievable optical depth. For fixed atomic density, adjusting the aspect ratio of the BEC to a cigar-shaped cloud with the long-axis aligned with the tightly focused probe beam increases the optical density. In this geometry, the limiting factor now becomes the Rayleigh length of the beam. Confining the atoms to a cylinder of diameter equal to the beam waste, and length equal to the Rayleigh length, and optimising the size of the beam waist gives a maximum optical depth of
\begin{equation}
d \leq \sigma_0\sqrt{\frac{N \rho}{\lambda}} \, ,
\end{equation}  
where $\rho$ is the atomic density, and $\lambda$ is the optical wavelength. For a $N=10^5$ $^{87}$Rb BEC using the D1 line, optimising the optical depth for a fixed atomic density of $10^{14}$/cm$^{3}$, gives an optical depth of $d=387$, and $\xi_\mathrm{opt} \approx 0.23$. In order to further improve the achievable squeezing, the effective optical depth, and therefore the achievable level of QND squeezing further, a high finesse optical cavity could be employed. 

\subsection{One-axis twisting.} 
\begin{figure}[h!]
    \includegraphics[width=\columnwidth]{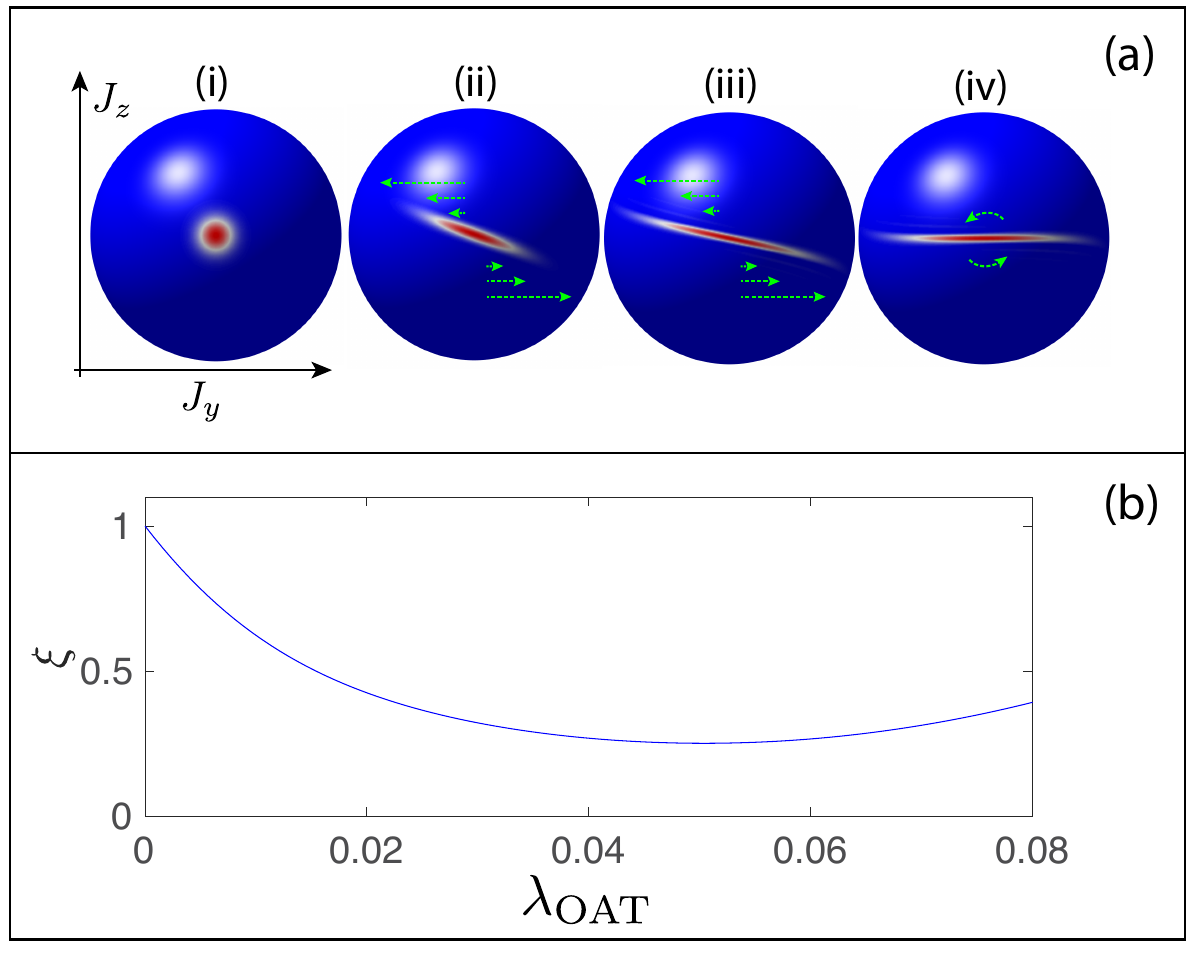}
    \caption{\label{fig:oat1} (a) Evolution of the Wigner quasi-probability distribution for an initial coherent spin state of 100 atoms under OAT dynamics, for (i) $\lambda_\mathrm{OAT} = 0$, (ii) $\lambda_\mathrm{OAT} = 0.025$, (iii) $\lambda_\mathrm{OAT} = 0.05$, (iv) $\lambda_\mathrm{OAT} = 0.05$, with an additional rotation around the $\Jhat_x$ axis applied to convert the squeezing into the $\Jhat_z$ direction. (b): Spin squeezing parameter as a function of $\lambda_\mathrm{OAT}$.  }
\end{figure}

One-axis twisting (OAT) dynamics is caused by a Hamiltonian of the form
\begin{eqnarray}
	\hat{H}_{\mathrm{OAT}} = \hbar \chi_{\mathrm{OAT}}(t) \hat{J}_z^2, \label{H_oat}
\end{eqnarray}
and leads to correlations between the relative number difference, and relative phase degrees of freedom \cite{Kitagawa:1993}. This results in a `shearing' of the quantum state on the Bloch sphere, and a narrowing of the spin distribution along one axis (figure \ref{fig:oat1}). In a two component BEC, OAT dynamics naturally arises from the inter-atomic interactions \cite{Sorensen:2001b, Li:2008, Esteve:2008, Li:2009, Haine:2009, Haine:2014, Laudat:2018}. Introducing the usual Bosonic field operators for hyperfine state $|j\rangle$, $\hat{\psi}_j(\boldr)$, which obey the usual commutation relations
\begin{equation}
\left[\hat{\psi}_i(\boldr) , \psihatd_j(\boldr^\prime)\right] = \delta_{ij}\delta(\boldr-\boldr^\prime) \, ,
\end{equation} 
the Hamiltonian term describing inter-atomic interactions is
\begin{align}
\hat{H}_\mathrm{int} = \sum_{i,j} \frac{U_{ij}}{2}\int \psihatd_i(\boldr)\psihatd_j(\boldr)\psihat_i(\boldr)\psihat_j(\boldr) \dr \, .
\end{align}
Making a single-mode approximation as in \cite{Haine:2014}, $\psihat_i(\boldr,t) \approx \ahat_i u_i(\boldr,t)$ and ignoring terms linear in $\Jhat_z$, we recover \eq{H_oat}, with $\chi(t) = 4(\chi_{11}(t) + \chi_{22}(t) - 2\chi_{12}(t))$, with 
\begin{equation}
\chi_{ij}(t) = \frac{U_{ij}}{2\hbar}\int |u_i(\boldr,t)|^2 |u_j(\boldr,t)|^2 \dr \, .
\end{equation}
The total effective OAT interaction is then given by the unitary $\hat{U}_\mathrm{OAT} = \exp(-i\lambda_\mathrm{OAT} \Jhat_z^2)$, where $\lambda_\mathrm{OAT} = \int_0^T \chi(t)dt$, and $T$ is the duration of the state preparation time. In the recent proposal by Szigeti \etal \cite{Szigeti:2020}, it was shown that spin-squeezing could be created by inducing significant OAT dynamics by spatially separating the two clouds in during the usual pre-expansion phase that usually proceeds atom interferometry. This sets $\chi_{12} \rightarrow 0$, significantly increasing $\chi$. As the clouds then expand, the magnitude of $\chi$ decreases to zero, causing $\lambda_\mathrm{OAT}$ to plateau. Benchmark calculations place the maximum interaction strength currently achievable at $\lambda_1 = 6.5 \times 10^{-5}$ for a system of $N = 10^5$ atoms \cite{Szigeti:2020}, which is considerably less than the optimum value for this number of atoms ($\lambda_\mathrm{opt} \approx 53 \times 10^{-5} $). However, our recent calculations indicate that interaction strengths of up to $\lambda_2 = 9.8\times 10^{-5}$ are possible by instantaneously applying an inwards focusing potential immediately before expansion, in a similar method to that proposed by \cite{Corgier:2021}. This value of $\lambda_\mathrm{OAT}$ is based on modeling the delta-kick scheme proposed by \cite{Corgier:2021} using the same simulation technique as \cite{Szigeti:2020}. These results are currently being prepared for publication. We use these values as references throughout our discussion.

\section{Simulating the hybrid method}\label{sec:hybrid}
In order to simulate the combination of QND and OAT dynamics, we cannot rely on simple models that give the spin-squeezing parameter for QND dynamics. This is because we need to know the form of the full quantum state, in order to perform the subsequent OAT dynamics. In particular, the magnitude of the anti-squeezing in the conjugate ($\hat{J}_y$) axis will have a significant effect on the subsequent OAT evolution. The truncated Wigner (TW) method \cite{Walls:2008} has been successfully used to simulate BEC dynamics \cite{Steel:1998, Sinatra:2002, Norrie:2006, Drummond:2017}. Importantly, the TW method can be used to model the production of non-classical correlations within the condensate \cite{Haine:2011, Opanchuk:2012, Ruostekoski:2013, Nolan:2016} including those generated OAT \cite{Haine:2009,Haine:2014, Haine:2018, Szigeti:2020} and atom-light interactions \cite{Haine:2016, Kritsotakis:2021}. The TW method also works well for large numbers of atoms, and can easily incorporate loss due to spontaneous emission. The derivation of the TW method has been described in detail elsewhere \cite{Drummond:1993, Steel:1998, Blakie:2008}. Briefly, the equation of motion for the Wigner function for the system can be found from the von-Neumann equation by using correspondences between differential operators on the Wigner function and the original quantum operators \cite{Gardiner:2004b}. By truncating third- and higher-order derivatives (the TW approximation), a Fokker-Planck equation (FPE) is obtained. The FPE is then mapped to a set of stochastic differential equations for complex variables $\{\alpha_1(t), \alpha_2(t), \beta(t)\}$, which loosely correspond to the annihilation operators of the system $\{\ahat_1(t), \ahat_2(t), \bhat(t)\}$, with initial conditions stochastically sampled from the appropriate Wigner distribution \cite{Blakie:2008, Olsen:2009}. Moments of observables are then calculated via the mapping $\langle \{ f((\ahat_1,\ahatd_1, \ahat_2,\ahatd_2, \bhat, \bhatd)\}_\mathrm{sym}\rangle \rightarrow \overline{f(\alpha_1, \alpha_1^*, \alpha_2, \alpha_2^*, \beta, \beta^*)}$, where `sym' denotes symmetric ordering and the overline denotes the mean over many stochastic trajectories.

\subsection{QND simulation}
We begin by simulating the QND dynamics of a series of light pulses, each of duration $t_p$, interacting with the BEC sequentially.  In the absence of loss due to spontaneous emission, mapping \eq{HQND} to the TW method, we find
\begin{subequations}
\begin{eqnarray}
i \ddt \alpha_1 &= \frac{\chi_{\mathrm{QND}}}{2} \abs{\beta_j}^2 \alpha_1 \\
i \ddt \alpha_2 &= \frac{\chi_{\mathrm{QND}}}{2} \abs{\beta_j}^2 \alpha_2 \\
i \ddt \beta_j &=  \chi_\mathrm{QND} \mathcal{J}_z \beta_j
\end{eqnarray}
\end{subequations}
where $\mathcal{J}_z = \frac{1}{2}(\abs{\alpha_1}^2 - \abs{\alpha_2}^2)$, and $\beta_j$ is the TW variable associated with the $j$th pulse. By assuming our pulses are short compared to the relevant atomic time-scales, we can take the continuum limit by introducing the parameter
\begin{equation}
\beta(t) = \frac{1}{\sqrt{t_p}}\sum_j \Pi_j(t)\beta_j
\end{equation}
where 
\begin{equation}
\Pi_j(t) =
    \begin{cases}
        1 & \text{if } jt_p < t \leq (j+1) t_p\\
        0 & \text{otherwise}.
    \end{cases}
\end{equation}
By taking the limit $t_p \rightarrow 0$, these equations can be solved analytically to give
\begin{subequations}
\begin{eqnarray}
\alpha_1(t) &= \exp\left(-i\frac{\lambda_{\mathrm{QND}}}{2}N_p(t) \right) \alpha_1(0) \\
\alpha_2(t) &= \exp\left(i\frac{\lambda_{\mathrm{QND}}}{2} N_p(t)\right) \alpha_2(0) \\
\beta(t) &=   \exp\left(-i \lambda_\mathrm{QND}\mathcal{J}_z \right) \beta(0)
\end{eqnarray}
\end{subequations}
where $N_p(t) = \int_0^t \abs{\beta(t^\prime)}^2 dt$, 
\begin{align}
\lambda_\mathrm{QND} = \chi_\mathrm{QND}t_p =  \frac{2\sigma_0 \Gamma}{A\Delta}.
\end{align}
and the initial condition are given by
\begin{subequations}
\begin{align}
\alpha_1(0) &= \sqrt{\frac{N_t}{2}} + \nu_1 \\
\alpha_2(0) &= \sqrt{\frac{N_t}{2}} + \nu_2 \\
\beta_{in}(t) &= \beta_0 + w(t)
\end{align}
\end{subequations}
where $\nu_j$ is complex Gaussian noise satisfying $\overline{\nu_i^*\nu_j} = \frac{\delta_{ij}}{2}$, and $\overline{w^*(t)w(t^\prime)} =\frac{1}{2}\delta(t-t^\prime)$.

We introduce the effect of spontaneous emission by noting that the fraction of atoms lost from each component in the duration of one pulse is $f = 1-e^{-\eta(t)}$, where
\begin{equation}
\eta(t) = \frac{2\sigma_0}{A} \left(\frac{\Gamma}{\Delta}\right)^2 |\beta_0|^2 t \,.
\end{equation}
By treating loss from the atomic system as an introduction of vacuum noise in the standard way \cite{Kritsotakis:2021}, we obtain 
\begin{subequations}
\begin{eqnarray}
\alpha_1(t) &=& \exp\left(-i\frac{\lambda_{\mathrm{QND}}}{2}N_p(t) \right) \alpha_1(0)\sqrt{1-f(t)}  \notag \\
&+& \sqrt{f(t)} v_1(0) \\
\alpha_2(t) &=& \exp\left(i\frac{\lambda_{\mathrm{QND}}}{2} N_p(t)\right) \alpha_2(0)\sqrt{1-f(t)}  \notag \\
&+& \sqrt{f(t)} v_2(0) \\
\beta_\mathrm{out}(t) &=&   \exp\left(-i \lambda_\mathrm{QND}\mathcal{J}_z(t) \right) \beta_\mathrm{in}(t)
\end{eqnarray}
\label{eqs_QND_full}
\end{subequations}
where $v_j(0)$ is complex gaussian noise satisfying $\overline{v_i^*v_j} = \frac{\delta_{ij}}{2}$. Here, we have defined $\beta_\mathrm{out}(t)$ as the light exiting the BEC after interacting with the atoms, and $\beta_\mathrm{in}(t)$ as the input light. 

Equations \ref{eqs_QND_full} create correlations between the relative population imbalance of the atoms, and the phase of the light. In order to convert this into spin-squeezing, the population imbalance is inferred from the phase-quadrature of the light, which is obtained via homodyne measurement, and is represented by the quantity 
\begin{equation}
Y = \frac{1}{\sqrt{T}}\int_0^Ti\left(\beta_\mathrm{out}(t) - \beta^*_\mathrm{out}(t)\right) dt \, .
\end{equation}
This information is then used to implement the feedback step, by rotating the atomic state around the $\Jhat_y$ axis by an angle $\theta_y$ proportional to the result of a measurement on the quadrature of the light. Specifically, we perform the transformation
\begin{subequations}
\begin{align}
\alpha_1 \rightarrow \cos \frac{\theta_y}{2} \alpha_1 + \sin \frac{\theta_y}{2} \alpha_2 \\
\alpha_2 \rightarrow \cos \frac{\theta_y}{2} \alpha_2 - \sin \frac{\theta_y}{2} \alpha_1 
\end{align}
\end{subequations}
where
\begin{equation}
\theta_y = \sin^{-1} \left(\frac{Y}{\lambda_\mathrm{QND} N_t \beta_0 T}\right) \, .
\end{equation}
The spin-squeezing parameter calculated via this method is shown in figure \ref{fig:QNDScaling}, and shows excellent agreement with the analytic solution. 

\subsection{Combining OAT and QND}
In order to simulate the hybrid scheme, we take the solution of  equations [\ref{eqs_QND_full}], perform a rotation by an amount $\theta_\mathrm{QND}$, and use this as the initial condition for the OAT dynamics. OAT dynamics is simulated in TW via mapping equation \ref{H_oat} to the set of ODEs for the TW variables 
\begin{subequations}
\begin{align}
i \ddt \alpha_1 = \frac{\chi(t)}{2}\left(\abs{\alpha_1}^2 - \abs{\alpha_2}^2\right)\alpha_1 \\
i \ddt \alpha_2 = -\frac{\chi(t)}{2}\left(\abs{\alpha_1}^2 - \abs{\alpha_2}^2\right)\alpha_2  
\end{align}
\end{subequations}
which has the simple analytic solution
\begin{subequations}
\begin{align}
\alpha_1(t) &= \exp\left(-i\frac{\lambda_\mathrm{OAT}(t)}{2}\left(\abs{\alpha_1}^2 - \abs{\alpha_2}^2\right)\right)\alpha_1(0) \\
\alpha_2(t) &= \exp\left(i\frac{\lambda_\mathrm{OAT}(t)}{2}\left(\abs{\alpha_1}^2 - \abs{\alpha_2}^2\right)\right)\alpha_2(0) 
\end{align}
\end{subequations}
After these dynamics, the state is rotated around the $J_x$ axis by an angle $\theta_\mathrm{OAT}$ in order to minimise the variance in $J_z$. 

\begin{figure}[h!]
    \includegraphics[width=\columnwidth]{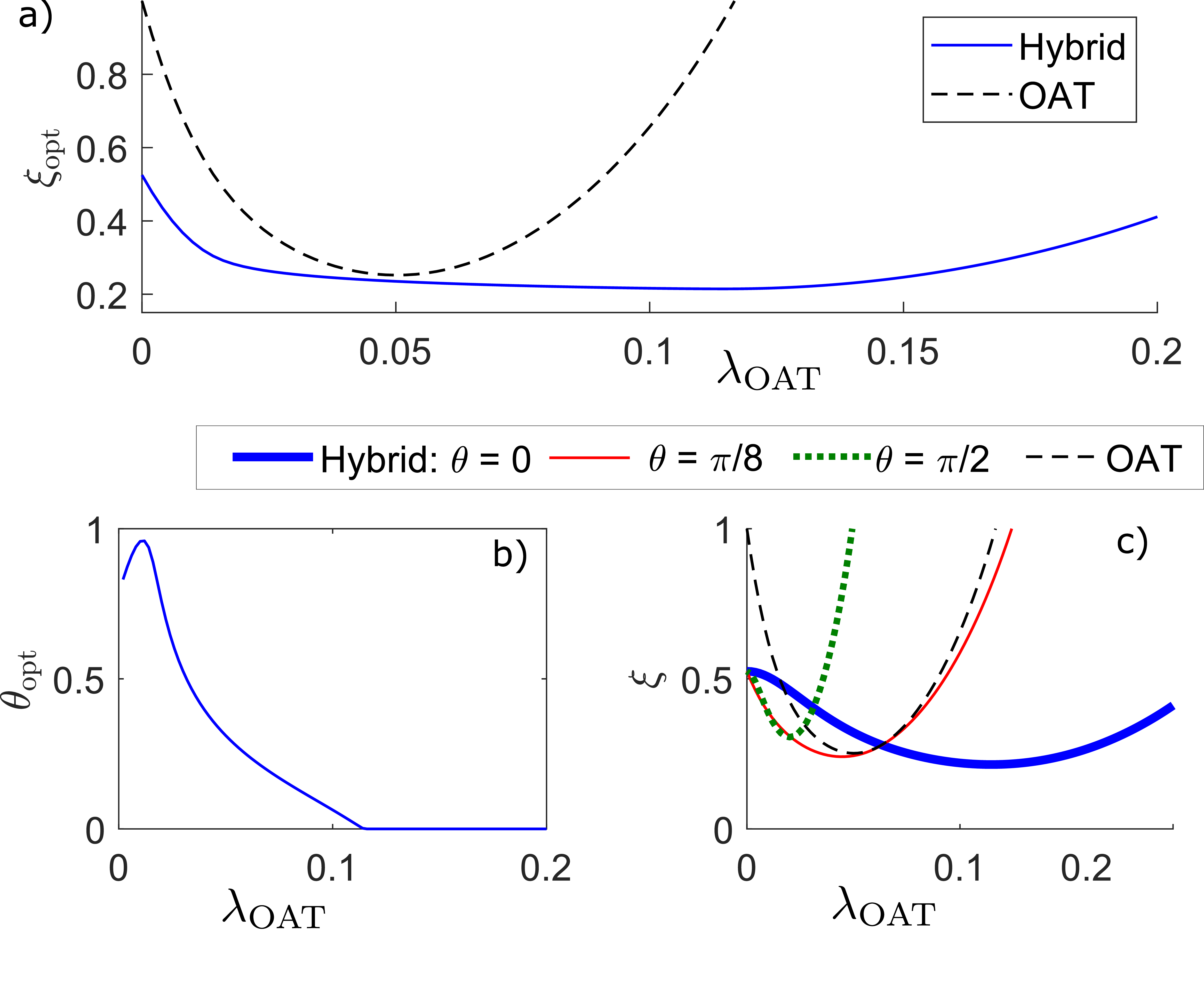}
    \caption{\label{fig:lossless} Comparison of OAT and the hybrid scheme in the absence of loss from QND, for $100$ atoms.  After QND, the state has a spin-squeezing parameter of $\xi = 0.5$. a) $\xi$ vs $\lambda_\mathrm{OAT}$ for three different values of $\theta_\mathrm{QND}$. Large angles increase the rate of spin-squeezing, but a worse optimum. b) Optimal value of $\theta_\mathrm{QND}$ as a function of $\lambda_\mathrm{OAT}$. Once $\lambda_\mathrm{OAT} \gg \lambda_\mathrm{opt}$, the optimum angle reduces to zero to capitalise on the increased spin-squeezing of states with small initial variance in $J_z$. c) Spin-squeezing parameter optimised over $\theta_\mathrm{QND}$. The hybrid model outperforms OAT over all interaction strengths. }
\end{figure}

To gain some intuition about the hybrid dynamics, we first simulate the system in the absence of spontaneous emission. Figure \ref{fig:lossless} shows how the spin-squeezing parameter evolves under OAT dynamics from an initially spin-squeezed state, compared to an initial CSS. Importantly, the role of the pre-OAT rotation by $\theta_\mathrm{QND}$ is clearly illustrated: for angles that increase the initial value of $V(\Jhat_z)$ to larger than a CSS, the OAT dynamics occurs much faster. When a smaller rotation angle is used, slightly better spin-squeezing is achieved, at the expense of much slower evolution.

\begin{figure}[h!] 
    \includegraphics[width=\columnwidth]{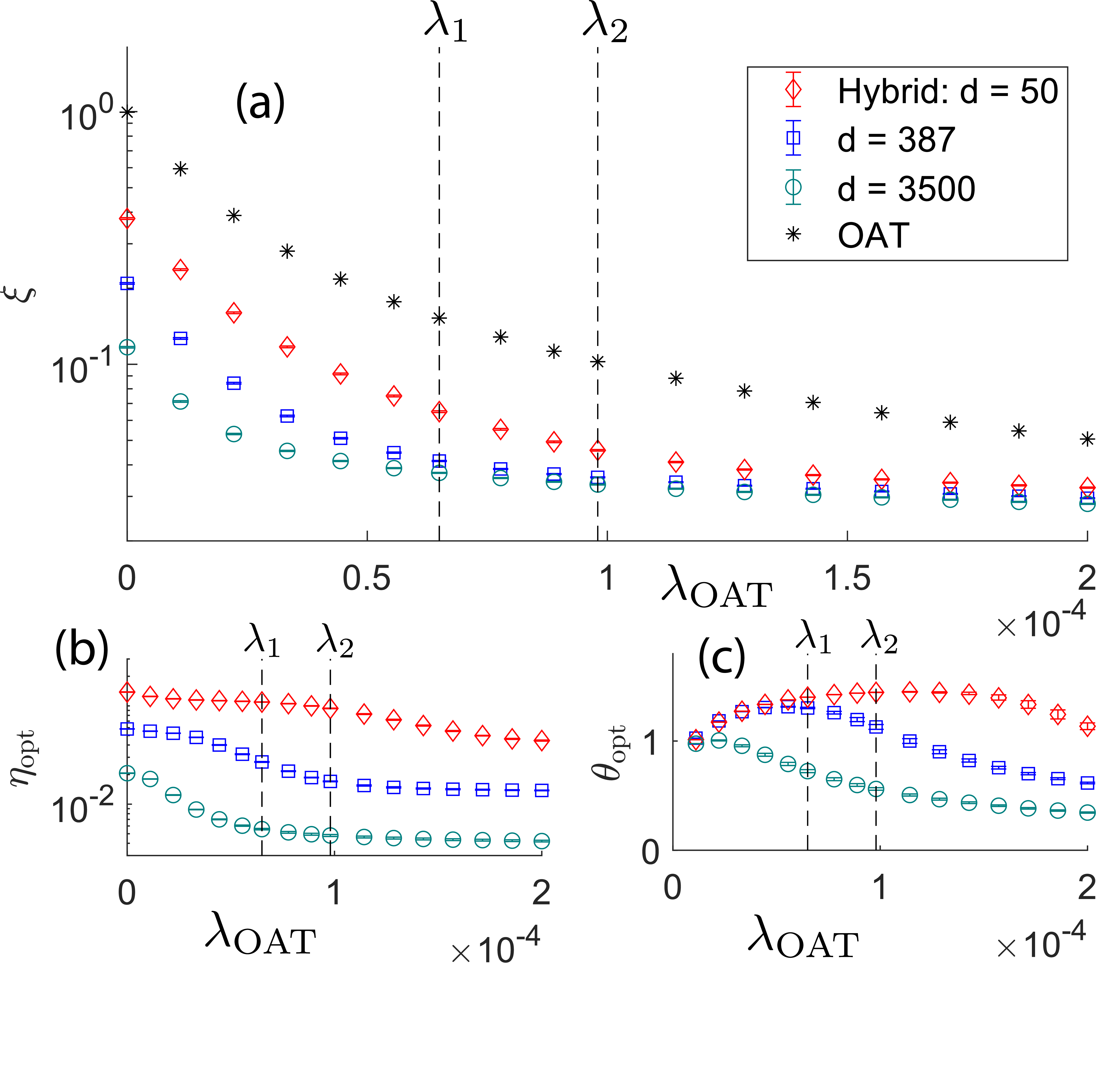}
    \caption{\label{fig:optimised} a) Squeezing parameter as a function of $\lambda_\mathrm{OAT}$ for the hybrid scheme compared to OAT, for three optical depths \(d= 50\) (red), 387 (blue) and 3500 (green). The hybrid method is optimised over $\eta$ and $\theta$. The dashed lines represent the OAT interaction strengths achievable via the scheme presented in \cite{Szigeti:2020}, ($\lambda_\mathrm{OAT} = \lambda_1$), and using a delta-kick scheme ($\lambda_\mathrm{OAT} = \lambda_2$). The hybrid method significantly outperforms the OAT scheme at both interaction strengths. b) Optimal measurement strength $\eta$ vs. $\lambda_\mathrm{OAT}$. $\eta_{\mathrm{opt}}$ decreases as $\lambda_\mathrm{OAT}$ increases as a consequence of OAT being more efficient for higher purity states. c) Optimal rotation angle vs. $\lambda_\mathrm{OAT}$. The behaviour of the rotation angle qualitatively mimics the lossless results.}
\end{figure}

When loss is included in the QND model, the final state after the QND interaction is no longer a minimum uncertainty state. The extra anti-squeezing in the $J_y$ direction will negatively affect the efficacy of the subsequent OAT dynamics. Figure \ref{fig:optimised} shows the spin-squeezing parameter including loss in the QND calculation, optimised over $\eta$ and $\theta_\mathrm{QND}$, compared to purely OAT dynamics, for realistic values of $\lambda_\mathrm{OAT}$, ie $\lambda_\mathrm{OAT} \le 2\times 10^{-4}$. We considered three optical depths corresponding to different exemplary QND laser configurations: $d = 50$, a poorly focused laser in free-space ($A = (1.5$mm$)^2$); $d = 387$, an optimally focused laser in free-space ($A = (0.53$mm$)^2$); and $d = 3500$, an optimally focused laser in a high-finesse optical cavity. with finesse $\mathcal{F} = 10^4$. The hybrid method outperformed OAT over all interaction strengths for each optical depth. For experimentally realisable interaction strengths, $\lambda_\mathrm{OAT} \le \lambda_1$, the high-finesse cavity achieved the most spin-squeezing by a significant margin. As the interaction strength increased, however, this advantage diminished. Specifically, at $\lambda_1$, the amount of spin-squeezing afforded by the optimal laser in free-space was on par with the high-finesse cavity, being roughly 4 times the spin-squeezing of OAT performed in isolation. Furthermore, at $\lambda_2$, all three optical depths were comparable, providing roughly 2.5 times the spin-squeezing of OAT. 

\begin{figure}[h!] 
    \includegraphics[width=\columnwidth]{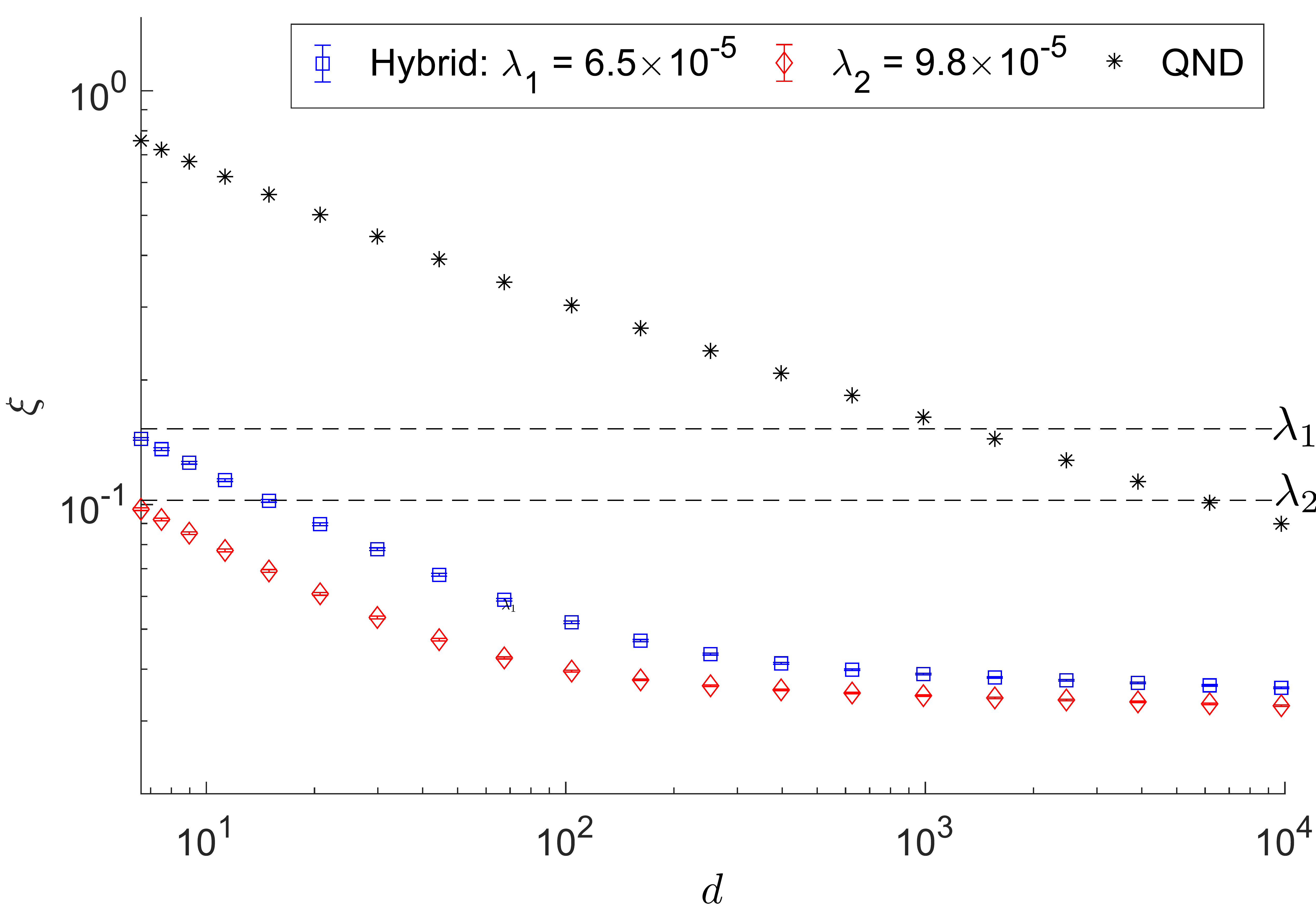}
    \caption{\label{fig:xivsd} Squeezing parameter scaling against resonant optical depth of the BEC for only QND (black), and the optimised hybrid method for a OAT interaction strength of \(\lambda_1\) (blue) and \(\lambda_2\) (red). Optical depths of \(d > 387\) require a cavity. The dashed lines represent OAT spin-squeezing at $\lambda_1$ and $\lambda_2$ respectively. QND at \(d = 10^4\) represents the best spin-squeezing available in experiments. The hybrid method exceeds this limit for \(d>10\) at both interaction strengths.}
\end{figure}

Figure \ref{fig:xivsd} shows the scaling of optimal QND spin-squeezing against different optical depths. For both interaction strengths, the hybrid model significantly outperformed QND. Specifically, at $d = 387$, the hybrid method spin-squeezed 5 times (5.8 times) more than optimal QND for total interaction $\lambda_1$ ($\lambda_2$). These results are unsurprising, as we expect OAT to enhance spin-squeezing of a QND SSS. However, the hybrid method is also seen to outperform QND in an ultra-high-finesse cavity ($d = 10^4$), the current leader in spin-squeezing demonstrated in proof-of-principle experiments \cite{Hosten:2016}. Specifically, the hybrid method outperformed in-cavity QND for optical depths of $d \approx 10$, meaning there is a large tolerance for imperfect laser focus. In addition, at the highest free-space optical depth,  $d = 387$, the hybrid method provides 2.5 times the amount of spin-squeezing over in-cavity QND. We therefore conclude that an experimental implementation of the hybrid method with modest requirements on size, weight, and power requirements could significantly increase spin-squeezing over QND.

\begin{figure}[h!] 
    \includegraphics[width=\columnwidth]{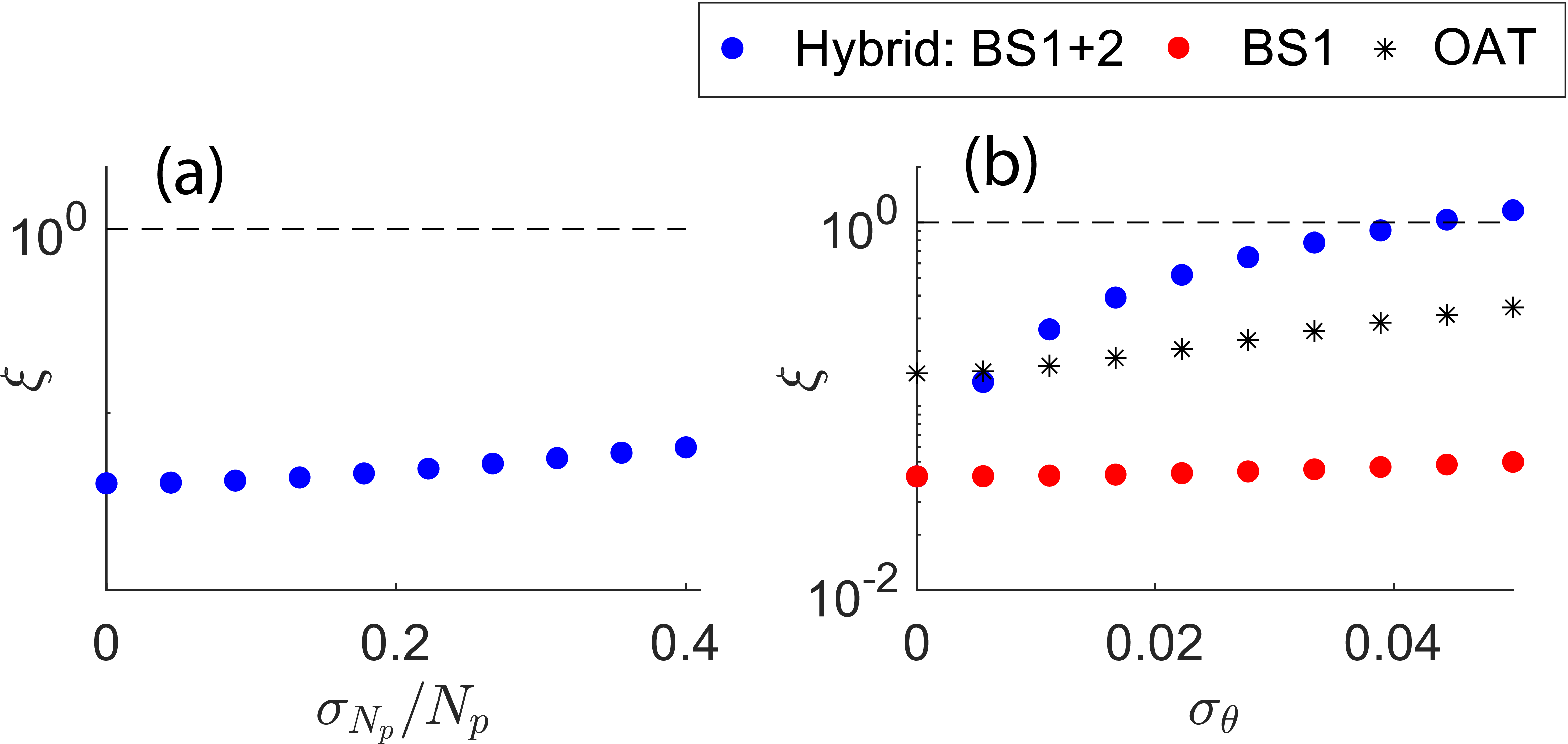}
    \caption{\label{fig:var} a) Squeezing parameter scaling against an introduced deviation in photon number (which controls the quantum non-demolition measurement strength) for the hybrid squeezing method. The parameter changes minimally for a deviation up to 30\% of the photon number. b) Squeezing parameter scaling against an introduced variance in rotation angle.}
\end{figure}

\section{Effect of experimental imperfections} While laser power may be stable over a single QND measurement, it may vary over multiple experiments. We incorporate these effects as fluctuations in the total photon during a QND measurement while fixing the beam-splitters before and after OAT is performed. Figure \ref{fig:var}a shows how the optimised hybrid method is affected by these changes for OAT interaction $\lambda_1 = 6.5\times 10^{-5}$. For fluctuations up to 30\% of the total photon number, there is virtually no change in spin-squeezing. Thus, the hybrid model is robust to fluctuations in the power of the laser used to perform QND.

We also investigate the effect of fluctuations in the power of the beam-splitter pulses. These will manifest as stochastic noise in the rotation angle of the state. Assuming only low-frequency noise, the angle will vary from shot-to-shot, rather than a single run, so we set the fluctuation in both beam-splitters to be the same. The impact on the optimised hybrid model for OAT interaction $\lambda_1$ is shown in Figure \ref{fig:var}b. Fluctuations of 0.05 radians in both beam-splitters is shown to completely wash out the spin-squeezing of the hybrid model (blue points). However, this is not an issue with the hybrid model, but rather of the stringent requirements for manipulating entangled states. Indeed, even for OAT performed on a CSS, much of the benefit is lost with fluctuations in the second beam-splitter (black asterisks). Furthermore, the hybrid scheme is seen to be robust to fluctuations in only the first beam-splitter up to 0.05 radians (red dots), indicating that the second beam-splitter is responsible for the degradation in squeezing. The hybrid method therefore has more stringent stability requirements than OAT and QND, but only due to complications with manipulating entangled states.

\section{Conclusion}
Our investigation of a hybrid method of QND and OAT indicates that significant gains to precision can be made with current state-of-the-art experiments without imposing high size, weight, and power requirements. We demonstrated that by first performing QND followed by OAT, the amount of spin-squeezing that can be achieved is greater than both OAT and in-cavity QND. As both of the techniques have been demonstrated in proof-of-principle experiments, the hybrid method can be incorporated with only small modifications to the experimental apparatus and procedure. Furthermore, we demonstrated that the hybrid method is robust to fluctuations in laser power when performing QND calculations, but is still subject to the stringent requirements of fluctuations in beam-splitter power for highly entangled states. The discussion in this paper has been limited to a single-mode model of OAT, which will need to be extended to a multimode model for more realistic predictions of spin-squeezing.  

This work has focussed on atomic interactions to achieve OAT dynamics. OAT dynamics can also be achieved by atom-light coupling via cavity feedback \cite{Schleier-Smith:2010b}, and has been used to create spin-squeezing \cite{Leroux:2010, Pedrozo-Penafiel:2020, Li:2022}. Interaction-based readouts \cite{Haine:2018b} have also been performed via this method \cite{Hosten:2016b}. As it is possible to smoothly transition between cavity-based OAT and QND dynamics \cite{Barberena:2023}, exploration of the hybrid method in these systems is an exciting direction for future research.

\begin{acknowledgments}
We would like to acknowledge fruitful discussions had with Stuart Szigeti, Zain Mehdi, and Karandeep Gill. We would also like to thank John Close for his input about experimental considerations for the model. SAH acknowledges support through an Australian Research Council Future Fellowship Grant No. FT210100809

\end{acknowledgments}

\bibliography{../../simon_bib.bib}

\end{document}